\documentclass[aps,prl,reprint,amsmath,nofootinbib]{revtex4-2}

\usepackage{graphicx}
\usepackage{bm}

\def\alphas{\alpha_{\rm s}}
\def\alphaqed{\alpha_{\scriptscriptstyle\rm EM}}
\def\Nc{N_{\rm c}}
\def\Nf{N_{\rm f}}
\def\lstop{\ell_{\rm stop}}
\def\eps{\epsilon}
\def\qhat{\hat q}
\def\qhatA{\hat q_{\rm A}}
\def\CA{C_{\rm A}}

\def\fac{{\rm fac}}
\def\LO{{\rm LO}}
\def\NLO{{\rm NLO}}
\def\net{{\rm net}}
\def\min{{\rm min}}
\def\eff{{\rm eff}}

\begin{document}

\title{Are gluon showers inside a quark-gluon plasma strongly coupled?
       \\a theorist's test}

\author{Peter Arnold}
\author{Omar Elgedawy}
\affiliation{%
    Department of Physics,
    University of Virginia,
    Charlottesville, Virginia 22904-4714, USA
}%
\author{Shahin Iqbal}
\affiliation{%
    National Centre for Physics,
    Quaid-i-Azam University Campus,
    Islamabad, 45320 Pakistan
}%

\date {\today}

\begin{abstract}%
We study whether in-medium showers of high-energy gluons can be treated
as a sequence of individual splitting processes $g{\to}gg$, or whether
there is significant quantum overlap between where one splitting
ends and the next begins.  Accounting for the Landau-Pomeranchuk-Migdal
(LPM) effect, we calculate such overlap effects to leading
order in high-energy $\alphas(\mu)$ for the simplest theoretical situation.
We investigate a measure of overlap effects that is independent
of physics that can be absorbed into an effective value $\hat q_{\rm eff}$
of the jet-quenching parameter $\hat q$.
\end{abstract}

\maketitle



\begin{figure}
\includegraphics[scale=0.7]{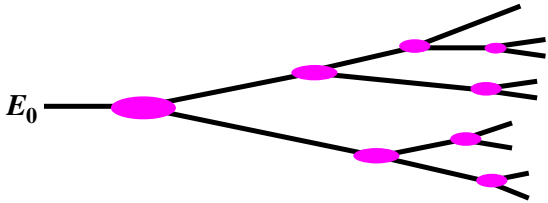}
\caption{
   \label{fig:shower}
   Schematic depiction of a high-energy shower in a medium.
   The splittings are nearly collinear, but tiny splitting angles
   have been exaggerated to make the drawing readable.
}
\end{figure}

When passing through matter,
very high energy particles lose energy by showering,
via the splitting processes of hard bremsstrahlung and pair production
induced by small-angle scatterings from the medium.
Fig.\ \ref{fig:shower} shows a cartoon of shower development, where
the energy $E_0$ of the initial high-energy particle is split among
more and more particles as time goes by, until eventually the remaining
particles have such low energy that they thermalize with the medium
(if the medium is thick enough to stop them before they leave).
We will focus on showers of very high energy
$(E \gg T)$ partons traversing a quark-gluon plasma of temperature $T$.
The quantum mechanical duration of a high-energy
splitting in the rest frame of the plasma
is known as the formation time.
We have drawn ovals in fig.\ \ref{fig:shower}
to represent the formation time (or
equivalently formation length) of each splitting, depicting
the formation times as small compared
to the time between splittings.  In that case, if one has results for
individual medium-induced splitting rates, one may statistically model shower
development by treating high-energy particles classically between splittings,
and rolling dice based on the splitting rates to decide when and how each
particle splits.  We call this a ``weakly-coupled'' picture of
in-medium shower development.

Alternatively, if formation times are large compared to
times between splittings, one may not treat different splittings as
quantum mechanically independent, and any classical picture of shower
development breaks down.  We will call that a ``strongly-coupled''
shower, which has been studied theoretically
for certain QCD-like theories (such as ${\cal N}{=}4$ supersymmetric QCD)
that can be studied with gauge-gravity duality
\cite{GubserGluon,HIM,CheslerQuark,adsjet12}.

As we will review, the distinction between
weakly- and strongly-coupled pictures of shower development is
controlled by the size of the running coupling $\alphas(\mu)$ at the
transverse momentum scale $\mu$ associated with high-energy splittings.
We will devise and calculate a theoretical measure of
how large $\alphas(\mu)$ can be before the weakly-coupled picture
of shower development breaks down.
Roughly, our approach will be to treat $\alphas(\mu)$ as
small but calculate the correction to the qualitative
picture of fig.\ \ref{fig:shower} by computing the correction from
{\it overlapping} formation times of two consecutive splittings.
We will have to carefully sharpen the question we ask
in order to factorize out effects of soft bremsstrahlung.
This paper aims to give a broad overview of our method and conclusion,
with many details and derivations left to a companion paper \cite{finale2}.

The formalism for making such calculations is challenging, and so
we take the simplest possible theoretical
situation.  (i) Imagine a quark-gluon plasma that is
static, homogeneous, and large enough to completely stop the shower.
(ii) Imagine that we start with a single high-energy parton
that is very close to on-shell.  (This ignores, for example,
the initial shower of decreasing virtuality that takes place
when a high-energy parton is
scattered out of a nucleon in a relativistic collision.)
(iii) Treat the elastic scattering of high-energy partons from the medium
in multiple-scattering ($\qhat$) approximation, which is that
the typical total transverse momentum change $p_\perp$ after traveling
through a length $L$ of medium behaves like a random walk,
$\langle p_\perp^2 \rangle = \qhat L$, where the proportionality
constant $\qhat$ is determined by the medium.
(iv) Take the large-$\Nc$ limit, where $\Nc$ is the number of
quark colors.
(v) Focus on gluon-initiated showers, and so the only
relevant splittings in fig.\ \ref{fig:shower}
are $g {\to} gg$ in the large-$\Nc$ limit.
All of these assumptions could in principle be relaxed in the formalism
that we use, but that would
make the calculations much more difficult.

Before proceeding, we review some parametric scales associated with
single splittings (such as $g {\to} gg$), shower development, and the
weakly-coupled picture of fig.\ \ref{fig:shower}.  Formation times
grow with energy.
At sufficiently high energy ($E \gg T$ in our case),
the formation time $t_{\rm form}$ of high-energy splittings
becomes large compared to the mean free
time $\tau_{\rm scatt}$ for elastic scattering from
the medium;
many scatterings take place during a single splitting, which causes
a very significant reduction of the splitting rate, known as the
Landau-Pomeranchuk-Migdal (LPM) effect \cite{LP1,LP2,LPenglish,Migdal}.
The treatment of the LPM effect in QCD was originally worked out by
Baier, Dokshitzer, Mueller, Peigne, and Schiff \cite{BDMPS1,BDMPS2,BDMPS3}
and by Zakharov \cite{Zakharov1,Zakharov2}
(BDMPS-Z).
In that limit, the formation time scales parametrically as
$t_{\rm form} \sim \sqrt{\omega/\qhat}$ in QCD, where $\omega \gg T$
is the energy of the least-energetic daughter of the splitting.
The typical scale $\mu$ of transverse momentum transferred from the
medium during the formation time is of order
\begin{equation}
  \mu \sim \sqrt{\qhat t_{\rm form}} \sim (\qhat \omega)^{1/4} .
\label{eq:mu}
\end{equation}
This is also the typical scale of the relative transverse momenta of
the two daughters of the splitting.

For simplicity, focus for now on
{\it democratic} splitting of a particle with energy $E$,
meaning that the two daughters have roughly comparable energies.
In the high-energy limit,
the probability of a democratic splitting is parametrically of order
$\alpha(\mu)$ per formation time, where $\alpha(\mu)$ is the running
QCD coupling.  Note that $\mu$ grows with energy $\omega \sim E/2$
in (\ref{eq:mu}).
Now consider two, consecutive,
democratic splittings.  Then the energies
and so formation lengths characteristic of the two consecutive splittings
are the same order of magnitude.
Since the probability of a splitting is parametrically $\alpha$
per formation time, the typical distance between splittings
will be of order $t_{\rm form}/\alpha$, and the probability of the
two consecutive splittings overlapping
will be order $\alpha$.  So, naively, the weak-coupling picture of showers
corresponds parametrically to $\alpha(\mu)$ small, and that picture
fails when $\alpha(\mu)$ is large.

That is a naive statement because the preceding argument was for
democratic splittings.  Refs.\ \cite{Blaizot,Iancu,Wu} have shown
that the probability of a hard splitting overlapping with {\it soft}
bremsstrahlung is enhanced by a large double logarithm in QCD,
similar to double logarithms in small-$x$ physics but with
some kinematic limits different.
They found that, even if $\alphas(\mu)$ is small,
such overlaps have large effects on energy loss when the double logarithm
compensates.  In our case of splitting of a high-energy particle of
energy $E$ in a thick quark-gluon plasma of temperature $T$,
``soft'' gluon energy $\omega$ means $T \ll \omega \ll E$,
which is the range that contributes to the double logarithm.
So overlap effects become significant when
$\alphas(\mu) \ln^2(E/T)$ is large, which can happen even if
$\alphas(\mu)$ is somewhat small.  But they also found that these
double log effects can be absorbed into a redefinition of the
medium parameter $\qhat$.  In our situation here, that means that
the potentially large effects of a soft gluon bremsstrahlung overlapping
a hard
splitting process can be absorbed into the original
LPM/BDMPS-Z calculation of the hard $g{\to}gg$ splitting rate by taking
$\qhat \to \qhat_\eff(E) = \qhat + \delta \qhat$ in that
calculation, where
$\delta\qhat(E) \sim \alphas \qhat \ln^2(E/T)$.
They also showed (following \cite{LMW}) how to resum leading logs
to all orders in $\alphas(\mu)$.


\section {Refining the question}

\begin{figure}
\includegraphics[scale=1.0]{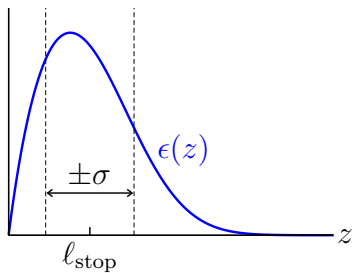}
\caption{
  \label{fig:eps}
  Energy deposition distribution $\eps(z)$.
}
\end{figure}

The goal of this paper is to construct a measure of the size of
overlap effects that {\it cannot}\/ be factorized away and absorbed
into an effective value for the medium parameter $\qhat$.  We start
with an idea proposed in ref.\ \cite{qedNfstop}.  For simplicity,
imagine for a moment a shower composed of democratic splittings.  The
distance between consecutive splittings is of order
$t_{\rm form}/\alpha \sim \alpha^{-1} \sqrt{E/\qhat}$, where the typical
energy $E$ of the individual shower particles decreases
rapidly as the shower develops.
A shower initiated by a single particle of energy $E_0$, moving in the
$z$ direction,  will therefore stop and
deposit all its energy into the medium in a distance of order
$\lstop \sim \alpha^{-1} \sqrt{E_0/\qhat}$, which depends on $\qhat$.
As a thought experiment, imagine measuring the distribution
$\eps(z)$ in $z$ of where that energy is deposited into the medium,
statistically averaged over many such showers.  (We do not track
the parametrically small
spread of the shower in the transverse directions.)
A qualitative picture is shown in
fig.\ \ref{fig:eps}.  We define $\lstop$ as the first moment
$\langle z \rangle \equiv E_0^{-1} \int dz\> z \, \eps(z)$
of this distribution.
Other features of
the distribution, such as its width
$\sigma = \sqrt{ \langle z^2 \rangle - \langle z \rangle^2}$,
are parametrically the same order as $\lstop \sim \alpha^{-1} \sqrt{E_0/\qhat}$.
Naively, the dependence on $\qhat$ would then {\it cancel}\/ in
a ratio such as $\sigma/\lstop$.
More generally, one may study any
aspect of what we will call the ``shape'' $S(Z)$ of the energy deposition
distribution $\eps(z)$.  By shape, we mean fig.\ \ref{fig:eps}
rescaled to units where $\lstop=1$ and normalized to have unit area
under the curve:
\begin{equation}
   S(Z) \equiv
   \frac{\langle z \rangle}{E_0} \, \eps\bigl( \langle z \rangle Z \bigr)
   ,
\end{equation}
where $Z \equiv z/\langle z \rangle$.
Naively, this shape function is insensitive to any physics
(such as soft bremsstrahlung) that can be
absorbed into the value of $\hat q$.

The shape $S(Z)$ and its moments
are insensitive to {\it constant}\/ shifts $\delta\hat q$ to $\hat q$.
However, the potentially
large double log correction,
arising from a soft bremsstrahlung overlapping a hard splitting,
is not constant: it depends logarithmically
on the energy scale $E$ of the underlying hard splitting.
So $\delta\hat q$ is different for different splittings in the
shower, and those differences do not exactly cancel in $S(Z)$.
As discussed in ref.\ \cite{qedNfstop} in the specific context of
$\sigma/\lstop$ (which is the second reduced moment of $S$),
the energy dependence
of the double-log
corrections from overlapping soft bremsstrahlung will lead to potentially
large single-log corrections
to the shape --- that is, corrections that are
$O\bigl( \alphas \ln(E_0/T) \bigr)$ instead of $O(\alphas)$.
The naive calculation of overlap corrections to $S(Z)$ will not
be completely independent of soft bremsstrahlung physics.

To proceed, consider a loose analogy with parton distribution functions (PDFs)
in the context of deep inelastic scattering (DIS) and other inclusive
processes.
The cross-sections factorize into (i) constituent cross-sections
of the partons and (ii) PDFs.  Beyond leading order (LO),
the constituent cross-sections
have initial-state collinear divergences in perturbation theory
that must be absorbed into the PDFs.  This requires introducing a
factorization scale $M_{\rm fac}$ to specify
exactly how much to absorb (analogous to
introducing the renormalization scale $\mu$ when absorbing ultraviolet
divergences) \cite{foot1}.
In next-to-leading order (NLO) perturbative calculations, the
answer depends on the choice of $M_\fac$, just as it depends on
the choice of renormalization scale $\mu$.  Theorists
typically set $M_\fac$ and $\mu$ to be the same and of order the
appropriate physics scale of the problem (e.g.\ $\sqrt{|Q^2|}$ in DIS)
in order to avoid large logarithms in the perturbative
expansion.  Typically, the exact choice of scale is varied
over a reasonable range to give a theory guess of uncertainty.  The higher
the order in perturbation theory, the less sensitive the result
to that variation.

We adopt a similar strategy.  We define $\qhat_\eff(\Lambda_\fac)$
to exactly
absorb all double and sub-leading single log behavior from overlapping
soft bremsstrahlung that has $\omega_{\rm soft} \le \Lambda_\fac$.
We will choose $\Lambda_\fac$ to be of order the relevant energy
scale of the problem (in the rest frame of the plasma).  Similar
to (\ref{eq:mu}), the corresponding
transverse momentum scale is $M_\fac \sim (\hat q \Lambda_\fac)^{1/4}$.
We will calculate all effects of overlapping
formation times on $S(Z)$ that have not already been absorbed into
$q_\eff(\Lambda_\fac)$.
Later, we will see that the question of
whether overlap effects that {\it cannot}\/ be absorbed into
$\hat q$ are large or small
is very insensitive to the exact choice of $\Lambda_\fac$.


\section {Shower Evolution}

The rates that contribute to LO+NLO, large-$\Nc$, gluon shower evolution
are called $[d\Gamma/dx]^\LO$,
$[\Delta\, d\Gamma/dx]^\NLO_{g\to gg}$, and
$[\Delta\, d\Gamma/dx\,dy]_{g\to ggg}$.
Formulas for these rates are given in refs.\ \cite{qcd,qcdI},
culminating the development of
refs.\ \cite{2brem,seq,dimreg,4point,QEDnf}.
Here, leading order refers to the LPM/BDMPS-Z rate
for a single,
{\it non}-overlapping splitting $g{\to}gg$, such as each individual splitting
shown in fig.\ \ref{fig:shower}.  Our ``LO'' rate encompasses an arbitrary
number of scatterings from the medium and does not assume that
the coupling $\alphas(T)$ of the quark-gluon plasma is perturbatively small.
Here, LO vs.\ NLO
refers only to how many powers of direct interactions $\alphas(\mu)$
between {\it high}-energy ($E \gg T$) partons are involved in the splitting.
$[d\Gamma/dx]^\LO$ is the differential LO rate for the energy to split
as $E \to xE + (1{-}x)E$.
$[\Delta \, d\Gamma/dx\,dy]_{g\to ggg}$ is a rate representing
the {\it overlap correction} to any two consecutive splittings,
$g{\to}gg{\to}ggg$, with final energy split as $E \to xE + yE + (1{-}x{-}y)E$.
(We also include $g\to ggg$ from direct
4-gluon vertices in $[\Delta \, d\Gamma/dx\,dy]_{g\to ggg}$ \cite{qcdI}.)
$[\Delta \, d\Gamma/dx]^\NLO_{g\to gg}$ gives related one-loop
corrections to single splitting, such as from $g \to gg \to ggg \to gg$.
These rates are designed so that one may evolve the shower using
classical statistics for an evolution that contains both
$1{\to}2$ splittings, with differential rate
$[d\Gamma/dx]_{1\to2} = [d\Gamma/dx]^\LO + [\Delta\, d\Gamma/dx]^\NLO_{g\to gg}$,
and
$1{\to}3$ splittings, with rate
$[\Delta \, d\Gamma/dx\,dy]_{g\to ggg}$.
The latter rate can sometimes be negative because it contains the overlap
{\it correction}, which can have either sign \cite{seq}.
Negative $[\Delta \, d\Gamma/dx\,dy]_{g\to ggg}$ will not cause
any problem for the NLO analysis in this paper.

In our convention, final-state identical particle symmetry factors are
not included in the differential rates above.  So, since
all our high-energy particles are gluons, the total splitting rate would
be formally (ignoring the fact that it is infrared divergent)
\begin{equation}
   \Gamma =
   \frac{1}{2!} \int_0^1 dx \> \left[ \frac{d\Gamma}{dx} \right]_{1\to2}
   +
   \frac{1}{3!} \int_0^1 dx \int_0^{1-x} dy
        \> \left[ \frac{d\Gamma}{dx\,dy} \right]_{1\to3} .
\end{equation}

When a shower involves more than just
$1{\to}2$ splitting processes,
the shower evolution equation can be neatly packaged in terms
of what we call the ``net'' rate $[d\Gamma/dx]_\net$ \cite{qcd}
for a splitting to produce
one daughter of energy $xE$ (plus any other daughters) from a parent
of energy $E$.  In the case of generic $1{\to}2$ and $1{\to}3$ splittings,
\begin{equation}
   \left[ \frac{d\Gamma}{dx} \right]_\net
   =
   \left[ \frac{d\Gamma}{dx} \right]_{1\to2}
   +
   \frac1{2!} \int_0^{1-x} dy \left[ \frac{d\Gamma}{dx\,dy} \right]_{1\to 3} .
\label{eq:dGnet}
\end{equation}
Note that the integral of $[d\Gamma/dx]_\net$ over $x$ is {\it not}\/
the total rate $\Gamma$.
Instead, there is a very useful alternative relation \cite{finale2}:
$ \Gamma = \int dx \> x \, [ d\Gamma/dx ]_\net$.
In terms of $[ d\Gamma/dx ]_\net$, the shower evolution equation is
\cite{qcd,finale2}
\begin{multline}
  \frac{\partial}{\partial t} n(\zeta,E_0,t)
  =
  \int_0^1 dx
    \biggl\{
      - x 
        \left[
          \frac{d\Gamma}{dx} (x,\zeta E_0)
        \right]_\net
        n(\zeta,E_0,t)
\\
      + \frac{\theta(x-\zeta)}{x}
        \left[
          \frac{d\Gamma}{dx} \bigl(x,\tfrac{\zeta}{x} E_0\bigr)
        \right]_\net
        n\bigl( \tfrac{\zeta}{x}, E_0, t \bigr)
    \biggr\}
,
\label{eq:Nevolve0}
\end{multline}
where
$n(\zeta,E_0,t)$ is the number density in $\zeta$
of gluons with energy $\zeta E_0$ at time $t$.
$[d\Gamma(x,E)/dx]_\net$ is the net splitting rate (\ref{eq:dGnet}),
and $\theta$ is the unit step function.

We have implicitly integrated over final (post-overlap)
transverse momenta both in our rate calculations and in
$n(\zeta,E_0,t)$, and chosen a $p_\perp$-insensitive test of
overlap effects, because implicit
$p_\perp$ integration drastically simplifies
the calculation of rates \cite{foot2}.

We want to factor out (and
absorb into $\hat q$) the
double and single logs arising from soft bremsstrahlung with energy
$\omega' \le \Lambda_\fac$, and so, at NLO, we use a factorized version of
the net rate in evolution equations like (\ref{eq:Nevolve0}).
In the multiple-scattering ($\hat q$) approximation we
have used, the net rate (\ref{eq:dGnet}) is double-log infrared divergent,
but the factorized net rate will not be.  The computations \cite{qcd,qcdI} of
splitting rates used a small infrared (IR) cut-off $\omega_\min$ on
soft gluon energy.
With that IR regulator, the {\it factorized}\/ rate is then
\begin{multline}
   \left[ \frac{d\Gamma}{dx} \right]_\net^\fac
   =
   \left[ \frac{d\Gamma}{dx} \right]_\net
   -
   \frac{\CA\alphas}{4\pi}
   \left[ \frac{d\Gamma}{dx} \right]^{\rm LO}
\\ \times
   \int_{\omega_\min}^{\Lambda_\fac} \frac{d\omega'}{\omega'}
   \Bigl\{
     \ln\Bigl( \frac{E}{\omega'} \Bigr)
     - \bar s(x)
   \Bigr\} ,
\label{eq:dGnetFac}
\end{multline}
where $\CA=\Nc$ is the adjoint Casimir, the integral of the first term
in braces produces a double logarithm, and the single
log coefficient $\bar s(x)$ is given explicitly in refs.\ \cite{logs,logs2}.
The combination (\ref{eq:dGnetFac}) is finite as $\omega_\min \to 0$ and
should be independent of the
details of the {\it actual} physics \cite{LMW,Jacopo} that cuts
off the double logarithm in the infrared.

The evolution equation (\ref{eq:Nevolve0}) can be simplified if the
(factorized) net rate scales with energy as exactly $E^{-1/2}$ for
fixed $x$.
This depends on the details of how one
chooses $\Lambda_\fac$.  One choice might be
(i) $\Lambda_\fac \propto E_0$, the energy of the entire shower.
Absorbing double logs into $\hat q$, the ``leading-order''
description would then use
$\hat q_\eff(E_0)$ for all splittings in the shower.  A more refined
choice would be (ii) $\Lambda_\fac \propto E$, and so use
$\hat q_\eff(E)$ for each splitting, adjusted
for the parent's energy $E$ of that particular splitting.  An even
more refined choice would be to recognize that the formation time and
transverse momentum kicks associated with a $g {\to} gg$ splitting
are determined (regarding the LPM effect) by the energy of the
softest daughter, and so take (iii)
$\Lambda_\fac \sim \operatorname{min}\bigl(xE,(1-x)E\bigr)$.
In case (i),
due to the mismatch of $\Lambda_\fac$ and the energy of
individual splittings,
a part of (\ref{eq:dGnetFac}) will scale like
$E^{-1/2} \ln^2(\Lambda_\fac/E)$, which does not allow simplification
of the evolution equation.  Both cases (ii) and (iii) avoid logarithmic
dependence on $E$.  Because (iii) is the most natural choice, we
stick to that here.  Specifically, we choose
$\Lambda_\fac \propto x(1-x)E$, which is a smooth function of $x$ with
the desired parametric behavior.

To simplify the shower evolution equation, scale
$E^{-1/2}$ out of the rate by rewriting
$d\Gamma(x,E) = E^{-1/2} \, d\tilde\Gamma(x)$,
$t = E_0^{1/2} \tilde t$, and
$n(\zeta,E_0,t) = \tilde n(\zeta,\tilde t)$.  Then
\begin{multline}
  \frac{\partial}{\partial\tilde t}\, \tilde n(\zeta,\tilde t)
  =
  \zeta^{-1/2}
  \int_0^1 dx
   \biggl[ \frac{d\tilde\Gamma}{dx} (x) \biggr]_\net^\fac
\\ \times
    \biggl\{
      - x\,
        \tilde n(\zeta,\tilde t)
      + \frac{\theta(x-\zeta)}{x^{1/2}}
        \tilde n\bigl( \tfrac{\zeta}{x}, \tilde t \bigr)
    \biggr\}
.
\label {eq:Nevolve1}
\end {multline}
An even simpler equation can be found for the (rescaled) energy deposition
distribution \cite{finale2,qedNfstop},
\begin{equation}
  \frac{\partial \tilde\eps(\tilde z)}{\partial\tilde z}
  =
  \int_0^1 dx \> x \biggl[\frac{d\tilde\Gamma}{dx}(x) \biggr]_\net^\fac
  \bigl\{ x^{-1/2}\,\tilde\eps(x^{-1/2} \tilde z)
          - \tilde\eps(\tilde z) \bigr\} ,
\label{eq:epseq}
\end{equation}
where $\tilde\eps(\tilde z) \equiv E_0^{-1/2} \eps(E_0^{1/2}\tilde z)$
is normalized
so that $\int_0^\infty d\tilde z \> \tilde\eps(\tilde z) = 1$.
Simpler yet, the {\it moments} of
this distribution are
given recursively in terms of integrals of the net
rate  \cite{finale2,qedNfstop}:
\begin{equation}
   \langle \tilde z^n \rangle
   = \frac{n \langle \tilde z^{n-1} \rangle}
          { \int_0^1 dx \> x(1-x^{n/2})
            \bigl[ \frac{d\tilde\Gamma}{dx} \bigr]_\net^\fac } \,.
\end{equation}


\section{Results and Conclusions}

We find that the width $\sigma_S = \sigma/\lstop$ of the shape distribution
$S(Z)$ is
\begin{equation}
   \frac{\sigma}{\lstop}  =
   \left[ \frac{\sigma}{\lstop} \right]^\LO_\eff
   ( 1 + \chi\alphas + \mbox{higher order} )
 \end{equation}
where the relative size of overlapping formation-time corrections
not absorbed into $\hat q_\eff$ is
\begin{equation}
  \chi\alphas = (-0.019 \pm 0.001 \ln\kappa) \, \CA\alphas(\mu)
\label{eq:chi}
\end{equation}
for $\Lambda_\fac = \kappa x(1{-}x)E$ and $\mu = (\qhatA \Lambda_\fac)^{1/4}$,
where our canonical choice is $\kappa=1$.
Even for $\Nc\alphas(\mu)=1$, (\ref{eq:chi}) is a tiny, few-percent effect
(for any reasonable choice of $\kappa$).

\begin{figure}
\includegraphics[scale=0.3]{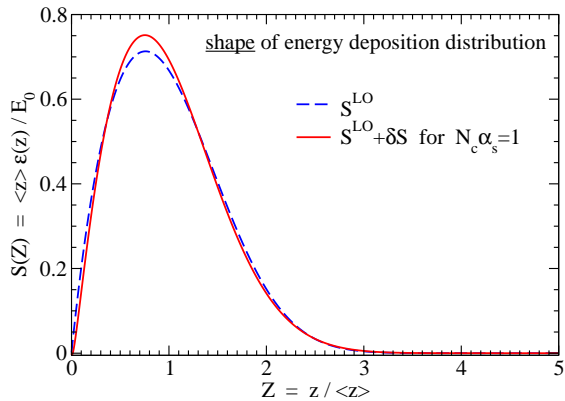}
\caption{
   \label{fig:S}
   Energy deposition shape with and without
   first-order overlapping formation time effects $\delta S$,
   for $\Nc\alphas=1$.
}
\end{figure}

Ref.\ \cite{finale2} gives some results for higher moments of
$S(Z)$ for gluon showers,
but it is more interesting to just look at how the function
$S(Z)$ itself changes.  Let $\delta S(Z)$ be the change in the shape function
to first order in overlap effects, i.e.\ to first order in $\alphas(\mu)$.
Fig.\ \ref{fig:S} depicts $S^\LO(Z)$ vs.
$S^\LO(Z) + \delta S(Z)$ for $\Nc\alphas=1$.  The difference is very
small and will be proportionally smaller for smaller $\Nc\alphas(\mu)$.

Though the corrections to the shape $S(Z)$ are very small for
large-$\Nc$ gluon showers (fig.\ \ref{fig:S}), the
corrections to quantities that {\it do} depend directly on $\hat q$ are
substantial, even when factorized.
The relative difference between $[d\Gamma/dx]_\net^\fac$ and $[d\Gamma/dx]^\LO$
can be of order $\Nc\alphas \times 100\%$ for democratic splittings
and is fairly sensitive to the choice of $\Lambda_\fac$ \cite{finale2}.

We should clarify that, when we use measurements of the shape
function to ``ignore all effects that can be absorbed into $\hat q$,'' we are
not claiming that those exact same effects also affect transverse momentum
broadening (the basis for our original definition of $\hat q$).
For our purpose here, think of $\hat q_{\rm eff}$
as an effective ``jet quenching'' parameter rather than a precisely
defined effective ``transverse momentum broadening'' parameter.
It is known that the
coefficient of the IR double logs are universal
in the sense that they affect both the same way \cite{Blaizot,Iancu,Wu}.
At least in the large-$\Nc$ limit, there is a
(more subtle) universailty for subleading, IR single logs as well
\cite{logs2}.
But we are unaware of any reason for such universality to hold beyond
logarithms.

In dramatic contrast to (\ref{eq:chi}),
ref.\ \cite{qedNfstop} analyzed $\sigma/\lstop$
for {\it charge} (rather than energy) deposition of
an {\it electron}-initiated shower in large-$\Nf$ QED, and the
analog of (\ref{eq:chi}) was found to be
$\chi\alphaqed = -0.87 \, \Nf \alphaqed(\mu)$ (and no factorization scale
need be introduced).  This is a large effect for $\Nf\alphaqed = 1$.
Ref.\ \cite{finale2} offers some crude, incomplete,
after-the-fact insight about the qualitative difference
with (\ref{eq:chi}) and motivates future study of (i) whether adding quarks
to our analysis would qualitatively change our conclusion and (ii) whether
overlap effects for
energy vs.\ charge stopping are qualitatively different.


\begin{acknowledgments}

This work supported, in part, by the U.S. Department
of Energy under Grant No.~DE-SC0007974.

\end{acknowledgments}



\begin{thebibliography}{}

\bibitem{GubserGluon}
  S.~S.~Gubser, D.~R.~Gulotta, S.~S.~Pufu and F.~D.~Rocha,
  ``Gluon energy loss in the gauge-string duality,''
  JHEP \textbf{10}, 052 (2008)
  [arXiv:0803.1470 [hep-th]].

\bibitem{HIM}
  Y.~Hatta, E.~Iancu and A.~H.~Mueller,
  ``Jet evolution in the ${\cal N}{=}4$ SYM plasma at strong coupling,''
  JHEP \textbf{05}, 037 (2008)
  [arXiv:0803.2481 [hep-th]].

\bibitem{CheslerQuark}
  P.~M.~Chesler, K.~Jensen, A.~Karch and L.~G.~Yaffe,
  ``Light quark energy loss in strongly-coupled ${\cal N}{=}4$
    supersymmetric Yang-Mills plasma,''
  Phys.\ Rev.\  D {\bf 79}, 125015 (2009)
  [arXiv:0810.1985 [hep-th]].

\bibitem{adsjet12}
  P.~Arnold, D.~Vaman,
  ``Jet quenching in hot strongly coupled gauge theories revisited:
  3-point correlators with gauge-gravity duality,''
  JHEP \textbf{10}, 099 (2010)
  [arXiv:1008.4023 [hep-th]];
  ``Jet quenching in hot strongly coupled gauge theories simplified,''
  JHEP \textbf{04}, 027 (2011)
  [arXiv:1101.2689 [hep-th]].

\bibitem{finale2}
  P.~Arnold, O.~Elgedawy and S.~Iqbal,
  ``The LPM effect in sequential bremsstrahlung: gluon shower development,''
  [arXiv:2302.10215 [hep-ph]].

\bibitem{LP1}
  L.~D.~Landau and I.~Pomeranchuk,
  ``Limits of applicability of the theory of bremsstrahlung electrons and
  pair production at high-energies,''
  Dokl.\ Akad.\ Nauk Ser.\ Fiz.\  {\bf 92} (1953) 535.

\bibitem{LP2}
  L.~D.~Landau and I.~Pomeranchuk,
  ``Electron cascade process at very high energies,''
  Dokl.\ Akad.\ Nauk Ser.\ Fiz.\  {\bf 92} (1953) 735.

\bibitem{LPenglish}
  L. Landau,
  {\sl The Collected Papers of L.D. Landau}\/
  (Pergamon Press, New York, 1965).

\bibitem{Migdal}
  A.~B.~Migdal,
  ``Bremsstrahlung and pair production in condensed media at high-energies,''
   Phys.\ Rev.\  {\bf 103}, 1811 (1956);

\bibitem{BDMPS1}
  R.~Baier, Y.~L.~Dokshitzer, A.~H.~Mueller, S.~Peigne and D.~Schiff,
  ``The Landau-Pomeranchuk-Migdal effect in QED,''
  Nucl.\ Phys.\  B {\bf 478}, 577 (1996)
  [arXiv:hep-ph/9604327];

\bibitem{BDMPS2}
  R.~Baier, Y.~L.~Dokshitzer, A.~H.~Mueller, S.~Peigne and D.~Schiff,
  ``Radiative energy loss of high-energy quarks and gluons in a
    finite volume quark - gluon plasma,''
  Nucl.\ Phys.\  B {\bf 483}, 291 (1997) [arXiv:hep-ph/9607355].

\bibitem{BDMPS3}
  R.~Baier, Y.~L.~Dokshitzer, A.~H.~Mueller, S.~Peigne and D.~Schiff,
  ``Radiative energy loss and $p_\perp$-broadening of high energy partons in
    nuclei,''
  {\it ibid.}\ {\bf 484} (1997)
  [arXiv:hep-ph/9608322].

\bibitem{Zakharov1}
 B.~G.~Zakharov,
 ``Fully quantum treatment of the Landau-Pomeranchuk-Migdal effect in
   QED and QCD,''
 JETP Lett.\  {\bf 63}, 952 (1996)
 [Pis'ma Zh.\ \'Eksp.\ Teor.\ Fiz.\  {\bf 63}, 906 (1996)]
 [arXiv:hep-ph/9607440].

\bibitem{Zakharov2}
 B.~G.~Zakharov,
 ``Radiative energy loss of high-energy quarks in finite size nuclear matter
   and quark-gluon plasma,''
 JETP Lett.\  {\bf 65}, 615 (1997)
 [Pis'ma Zh.\ \'Eksp.\ Teor.\ Fiz.\  {\bf 65}, 585 (1997)]
 [arXiv:hep-ph/9704255].

\bibitem{Blaizot}
  J.~P.~Blaizot and Y.~Mehtar-Tani,
  ``Renormalization of the jet-quenching parameter,''
  Nucl.\ Phys.\ A {\bf 929}, 202 (2014)
  [arXiv:1403.2323 [hep-ph]].

\bibitem{Iancu}
  E.~Iancu,
  ``The non-linear evolution of jet quenching,''
  JHEP \textbf{10}, 95 (2014)
  [arXiv:1403.1996 [hep-ph]].

\bibitem{Wu}
  B.~Wu,
  ``Radiative energy loss and radiative $p_{\bot}$-broadening of
    high-energy partons in QCD matter,''
  JHEP \textbf{12}, 081 (2014)
  [arXiv:1408.5459 [hep-ph]].

\bibitem{LMW}
  T.~Liou, A.~H.~Mueller and B.~Wu,
  ``Radiative $p_\bot$-broadening of high-energy quarks and gluons in
    QCD matter,''
  Nucl.\ Phys.\ A {\bf 916}, 102 (2013)
  [arXiv:1304.7677 [hep-ph]].

\bibitem{qedNfstop}
  P.~Arnold, S.~Iqbal and T.~Rase,
  ``Strong- vs. weak-coupling pictures of jet quenching: a dry run using QED,''
  JHEP \textbf{05}, 004 (2019)
  [arXiv:1810.06578 [hep-ph]].

\bibitem{foot1}
  See, e.g., chapter 7 of ref.\ \protect\cite{ellis}.

\bibitem{ellis}
  R.~K.~Ellis, W.~J.~Stirling and B.~R.~Webber,
  {\it QCD and collider physics},
  Cambridge University Press, 1996
  [Camb.\ Monogr.\ Part.\ Phys.\ Nucl.\ Phys.\ Cosmol.\ \textbf{8}].

\bibitem{qcd}
  P.~Arnold, T.~Gorda and S.~Iqbal,
  ``The LPM effect in sequential bremsstrahlung:
    nearly complete results for QCD,''
  JHEP \textbf{11}, 053 (2020)
  [{\it erratum} JHEP \textbf{05}, 114 (2022)]
  [arXiv:2007.15018 [hep-ph]].

\bibitem{qcdI}
  P.~Arnold, T.~Gorda and S.~Iqbal,
  ``The LPM effect in sequential bremsstrahlung:
    incorporation of ''instantaneous'' interactions for QCD,''
  JHEP \textbf{11}, 130 (2022)
  [arXiv:2209.03971 [hep-ph]].

\bibitem{2brem}
  P.~Arnold and S.~Iqbal,
  ``The LPM effect in sequential bremsstrahlung,''
  JHEP \textbf{04}, 070 (2015)
  [{\it erratum} JHEP \textbf{09}, 072 (2016)]
  [arXiv:1501.04964 [hep-ph]].

\bibitem{seq}
  P.~Arnold, H.~C.~Chang and S.~Iqbal,
  ``The LPM effect in sequential bremsstrahlung 2: factorization,''
  JHEP \textbf{09}, 078 (2016)
  [arXiv:1605.07624 [hep-ph]].

\bibitem{dimreg}
  P.~Arnold, H.~C.~Chang and S.~Iqbal,
  ``The LPM effect in sequential bremsstrahlung: dimensional regularization,''
  JHEP \textbf{10}, 100 (2016)
  [arXiv:1606.08853 [hep-ph]].

\bibitem{4point}
  P.~Arnold, H.~C.~Chang and S.~Iqbal,
  ``The LPM effect in sequential bremsstrahlung: 4-gluon vertices,''
  JHEP \textbf{10}, 124 (2016)
  [arXiv:1608.05718 [hep-ph]].

\bibitem{QEDnf}
  P.~Arnold and S.~Iqbal,
  ``In-medium loop corrections and longitudinally polarized gauge bosons
    in high-energy showers,''
  JHEP \textbf{12}, 120 (2018)
  [arXiv:1806.08796 [hep-ph]].

\bibitem{foot2}
  See, e.g., section 4.1.1 of ref.\ \cite{2brem}.

\bibitem{logs2}
  P.~Arnold,
  ``Universality (beyond leading log) of soft radiative corrections
    to $ \hat{q} $ in p$_\perp$ broadening and energy loss,''
  JHEP \textbf{03}, 134 (2022)
  [arXiv:2111.05348 [hep-ph]].

\bibitem{logs}
  P.~Arnold, T.~Gorda and S.~Iqbal,
  ``The LPM effect in sequential bremsstrahlung:
    analytic results for sub-leading (single) logarithms,''
  JHEP \textbf{04}, 085 (2022)
  [arXiv:2112.05161 [hep-ph]].

\bibitem{Jacopo}
  J.~Ghiglieri and E.~Weitz,
  ``Classical vs quantum corrections to jet broadening in a
    weakly-coupled Quark-Gluon Plasma,''
  JHEP \textbf{11}, 068 (2022)
  [arXiv:2207.08842 [hep-ph]].

\end{thebibliography}
\end{document}